\documentclass[pra,aps,superscriptaddress,twocolumn]{revtex4}

\usepackage{graphicx}

\usepackage{amsmath}
\usepackage{bm,CJK}
\usepackage{mathrsfs}
\usepackage{amsfonts}
\usepackage{mathtools}
\usepackage{graphicx}
\usepackage{subfigure}
\usepackage{braket}

\def\be{\begin{equation}}
\def\ee{\end{equation}}
\def\ba{\begin{eqnarray}}
\def\ea{\end{eqnarray}}

\begin{document}

\title{Quantum Algorithm for a Set of Quantum 2-SAT Problems}
\begin{CJK}{UTF8}{gbsn}
\author{Yanglin Hu(胡杨林)}
\affiliation{International Center for Quantum Materials, School of Physics, Peking University, 100871, Beijing, China}

\author{Zhelun Zhang(张哲伦)}
\affiliation{International Center for Quantum Materials, School of Physics, Peking University, 100871, Beijing, China}

\author{Biao Wu(吴飙)}
\email{wubiao@pku.edu.cn}
\affiliation{International Center for Quantum Materials, School of Physics, Peking University, 100871, Beijing, China}
\affiliation{Wilczek Quantum Center, School of Physics and Astronomy, Shanghai Jiao Tong University, Shanghai 200240, China}
\affiliation{Collaborative Innovation Center of Quantum Matter, Beijing 100871,  China}

\date{\today}
\begin{abstract}
We present a quantum adiabatic algorithm for  a set of quantum 2-satisfiability (Q2SAT) problem, 
which is a generalization of 2-satisfiability (2SAT) problem.
For a Q2SAT problem, we construct the Hamiltonian which is similar to that of a Heisenberg chain. All the solutions of the given Q2SAT problem span the subspace of the degenerate ground states. The Hamiltonian is adiabatically evolved so that the system stays in the degenerate subspace.  Our numerical results suggest that the time complexity of our algorithm is $O(n^{3.9})$ for yielding non-trivial solutions for problems with the number of clauses $m=dn(n-1)/2$ ($d\lesssim 0.1$). 
We discuss the advantages of our algorithm over the known quantum 
and classical  algorithms. 

~

~
\end{abstract}

\maketitle
\end{CJK}

\section{\label{sec:introduction}Introduction}

In 1990s, several quantum algorithms such as Shor's algorithm for factorization and Grover's algorithm for search \cite{nielsen_chuang_2010} were found to  have a lower time complexity than their classical counterparts. These quantum algorithms are based on discrete quantum operations, and are called quantum circuit algorithms. 

Quantum algorithms of a different kind were proposed by Farhi et al. \cite{Farhi1998,farhi_goldstone_gutmann_sipser}. In these algorithms, Hamiltonians 
are constructed for a given problem and the qubits are prepared initially in an easy-to-prepare state. The state of the qubits is then driven dynamically and 
continuously by the Hamiltonians and finally arrives at the solution state. Although quantum algorithms with Hamiltonians have been shown  
to be no slower than  quantum circuit algorithms~\cite{aharonov_dam_kempe_landau_loyd_regev,Yu_Huang_Wu},  they  have   found very limited success. In fact, due to exponentially small energy gaps~\cite{Boris}, they often can not even outperform classical algorithms. 
The random search problem is a rare exception, for which 
three different quantum Hamiltonian algorithms were proposed  and they can outperform classical algorithms. But still 
these Hamiltonian algorithms are just as fast as Grover's~\cite{Farhi1998,Cerf2000,Dam1,Wilczek_2020}.

Recently, quantum Hamiltonian algorithms were found for a different problem, independent sets of a graph~\cite{Wu_Yu_Wilczek,YWW} and
they can outperform their classical counterparts significantly. In this work, we apply it to  a set of quantum 2-satisfiability (Q2SAT) problems, which have two groups of solutions in the form of product states and entangled states. We aim to find solutions in the form of entangled states. For a given Q2SAT problem, we construct a Hamiltonian whose ground states are all the solutions of the problem. 
Initially we prepare the system in a trivial product solution state, we then evolve it in the subspace of degenerate ground states by slowly changing 
Hamiltonian parameters along a closed path. In the end we get a superposition of different solutions.  
Numerical calculation shows that  the time complexity of our quantum algorithm is $O(n^{3.9})$ for problems with $m=dn(n-1)/2$ ($d\lesssim 0.1$). $m$ is the number of clauses. There is a classical algorithm 
for the Q2SAT problem. Although its time complexity is better, it tends to find trivial product  solutions~\cite{Beaudrap_Gharibian,Arad_Santha_Sundaram_Zhang}. 
The quantum algorithm in Ref.~\cite{Farhi_Kimmel_Temme}
can find entangled solutions but  with a slower time complexity of $O(mn^2/\delta(n))$, where the energy gap 
$\delta(n)$ may be in the form of $n^{-g}$ (g positive).

\section{\label{sec:2-QSAT_problem_and_its_previous_algorithm}Quantum 2-Satisfiability Problem}
Quantum 2-satisfiability (Q2SAT) problem is a generalization of the well known 2-satisfiability (2SAT) problem~\cite{Beaudrap_Gharibian}. The algorithm for 2SAT problem is widely used in scheduling and gaming~\cite{Even_Itai_Shamir}. Besides, 2SAT problem is a subset of k-satisfiability problem (kSAT). Since 2SAT problem is a P problem while kSAT problem is a NP complete problem, kSAT problem has a great importance in answering whether P=NP. Similarly, Q2SAT problem is a subset of quantum k-satisfiability problem (QkSAT). It is expected that QkSAT problem is more complex than kSAT problem, and that quantum algorithms perform better than classical algorithms in QkSAT problem. Therefore, QkSAT problem could become a breakthrough in answering whether P=BQP and BQP=NP~\cite{Sergey}. 

In a 2SAT problem, there are $n$ Boolean variables and $m$ clauses. Each clause of two Boolean variables bans one of the four possible assignments. 
For example, the clause $(\neg x_i\vee x_j)$ bans the assignment $(x_i, x_j)=(1, 0)$. The problem 
is to find an assignment for all the variables so that all the clauses are satisfied.  For quantum generalization, we replace  the boolean variables with qubits and 
the clauses with two-qubit projection operators.  In a Q2SAT problem of $n$ qubits and $m$ two-qubit projection operators $\{\Pi_1,\Pi_2, ..., \Pi_m\}$,  the aim is to 
find a state $|\psi\rangle$ such that projections of the state are zeros, i.e., 
\begin{equation}
    \Pi_j |\psi\rangle = 0,\ \forall j\in{1,2,\cdots,m}\,.
\end{equation}
When all the projection operators project onto product states, Q2SAT problems go back to 2SAT problems. 

In this work we focus on a class of 2-QSAT problems, where all the projection operators are of an identical form 
\begin{equation}
\label{proj}
    \Pi_j = |\Phi_j\rangle\langle\Phi_j|,\ |\Phi_j\rangle = \alpha |1_{a_j}0_{b_j}\rangle + \beta |0_{a_j}1_{b_j}\rangle, \forall j\,,
\end{equation}
where $|\alpha|^2+|\beta|^2=1$ and $a_j,b_j$ label the two qubits acted on by $\Pi_j$.
This is a special case of the restricted Q2SAT problems discussed by Farhi et al \cite{Farhi_Kimmel_Temme}, i.e. 
where all the clauses are the same. 
These Q2SAT problems have apparently have two solutions, $|\psi\rangle=|0\rangle...|0\rangle$ and $|\psi\rangle=|1\rangle...|1\rangle$, 
which are product states. We call them trivial solutions. We are interested in finding non-trivial solutions which are entangled. 

A Q2SAT problem of $n$ qubits and $m$ two-qubit projections can be also viewed as a generalization of a graph with  $n$ vertices  and $m$ edges. As a result, in this work, we often refer to Q2AT problem as graph.

\section{\label{sec:pre_algorithm}Previous algorithms}
There are now several algorithms for Q2SAT problems. The algorithm proposed by Beaudrap et al in \cite{Beaudrap_Gharibian} and Arad et al in \cite{Arad_Santha_Sundaram_Zhang} is classical. The classical algorithm relies on that for every Q2SAT problem which has solutions, there is a solution that is the tensor product of one-qubit and two-qubit states, 
\begin{equation}
    |\psi_1\rangle = \prod_{r} |\psi_r\rangle \otimes \prod_{pq} |\psi_{pq}\rangle.
    \label{ps}
\end{equation}
where $|\psi_r\rangle$ is the state of the qubit $r$, $|\psi_{pq}\rangle$ is an entangled state of qubit $p$ and $q$, 
and the indices $r$ and $p,q$ do not overlap. This conclusion is drawn with the following proven fact. 
If a projection operator $\Pi_j$ projects onto an entangled state of qubits $a_j$ and $b_j$, then the solution has either of the following two forms: 
\begin{equation}
    |\psi_1\rangle=|\psi_{a_j b_j}\rangle \otimes |\mathrm{rest}\rangle,
\end{equation}
where $|\psi_{a_j b_j}\rangle$ is an entangled state of qubits $a_j$ and $b_j$, and
\begin{equation}
    |\psi_1\rangle=|\psi_{a_j}\rangle \otimes |\psi_{b_j}\rangle \otimes |\mathrm{rest}\rangle.
\end{equation}
where $|\psi_{a_j}\rangle$ and $|\psi_{b_j}\rangle$ are single-qubit states. Based on this feature, we conclude that a qubit involved in only one projection operator has entanglement with the other qubit of this projection operator, and that a qubit involved in more than one projection operators has no entanglement with other qubits. To find a solution of the form in Eq.(\ref{ps}), 
one can use the strategy of Davis-Putman's algorithm for 2-SAT problem. That is, we assign an initial state to a qubit, "propagate" the state to its adjacent qubits along projection operators, and finally find out the solution of this form. The above algorithm has a time complexity of $O(n+m)$, but it is impossible  to find a  solution where three or more qubits are entangled.

In the quantum algorithm in \cite{Farhi_Kimmel_Temme}, Farhi et al constructed a Hamiltonian 
\begin{equation}\label{eqn:Farhi_H}
    H=\sum_j\Pi_j,
\end{equation}
where $\Pi_j$ is of the form in Eq.(\ref{proj}). There is one-to-one correspondence between the
solutions of a Q2SAT problem and the ground states of its corresponding Hamiltonian. To see this, 
we consider a state $|\psi_2\rangle$. If a state $|\psi_2\rangle$ is a solution of the Q2SAT problem, then
\begin{equation}\label{eqn:ground}
    H|\psi_2\rangle =\sum_j \Pi_j |\psi_2\rangle =0,
\end{equation}
and if $|\psi_2\rangle$ is not a solution, then
\begin{equation}\label{eqn:excited}
    \langle \psi_2|H|\psi_2\rangle = \sum_j \langle \psi_2|\Pi_j |\psi_2\rangle >0.
\end{equation}
Therefore, $|\psi_2\rangle$ is a solution of the Q2SAT problem if and only if it is a ground state of the Hamiltonian $H$. The state is initialized to
\begin{equation}
    \rho_0=\frac{1}{2^n}\mathbb{I}.
\end{equation}
In each step of the algorithm, a projection operator $\Pi_j$ is selected and measured at random. If the result is $0$, then do nothing, otherwise a Haar random unitary transfromation is applied
\begin{equation}
    \Lambda_{a_j}(\rho)=\int d[U_{a_j}] U_{a_j}\rho U_{a_j},
\end{equation}
on one qubit $a_j$ of the two qubits $a_j$ and $b_j$ involved in $\Pi_j$. That is, the operation on the state in each step is 
\begin{equation}
    \mathscr{T}(\rho)=\frac{1}{m} \sum_{j} \mathscr{T}_j(\rho),
\end{equation}
where 
\begin{equation}
    \mathscr{T}_j(\rho)=(1-\Pi_j)\rho(1-\Pi_j)+\frac{1}{2}\sum_{a_j,b_j}\Lambda_{a_j,b_j}(\Pi_j \rho \Pi_j).
\end{equation}
Now set 
\begin{equation}
    T=\max \{\frac{49m^2 n^2}{2c^2},\frac{3mn^2}{2\epsilon}\},
\end{equation}
where $c(n)$ is the energy of the ground state and $\epsilon(n)$ is the energy gap between the ground state and the first excited state. It is assumed that  $\epsilon(n)\approx n^{-g}$ (g positive). After  steps of length $T$, 
the algorithm has a probability of at least $2/3$ to produce a state $\rho_T$ whose fidelity with 
the solution is at least $2/3$. The quantum algorithm has a time complexity of at least $O(mn^2/\epsilon)$, 
and gives a non-trivial solution.

\section{\label{sec:our_algorithm}Our algorithm}
Our algortihm follows the one proposed in Ref.\cite{Wu_Yu_Wilczek}. 
For a  Q2SAT problem of $n$ qubits and $m$ two-qubit projection operators $\{\Pi_1, \Pi_2,..., \Pi_m\}$, we construct a Hamiltonian similar to Eq. (\ref{eqn:Farhi_H})
\begin{equation}
    H_0= \Delta \sum_{j=1}^{m} \Pi_j \,,
\end{equation}
where $\Delta$ is a positive real number  and $\Pi_j$ is of the form in Eq.(\ref{proj}). Due to equations similar to Eq. (\ref{eqn:ground}) and Eq.(\ref{eqn:excited}), solutions of the problem have one-to-one correspondence to the ground states of $H_0$. 
The above Hamiltonian can be re-written in terms of  spin-1/2 operators $s^x,s^y,s^z$ as
\begin{eqnarray}
    H_0=&&\Delta \sum_{j=1}^m\Big\{ -s^z_{a_j} s^z_{b_j}-\frac{1}{2}(1-2|\beta|^2)(s^z_{a_j}-s^z_{b_j})\nonumber\\
    &&+2\mathrm{Re}(\beta)\sqrt{1-|\beta|^2}(s^x_{a_j}s^x_{b_j}+s^y_{a_j}s^y_{b_j})\nonumber \\
    &&+2\mathrm{Im}(\beta)\sqrt{1-|\beta|^2}(s^x_{a_j}s^y_{b_j}-s^y_{a_j}s^x_{b_j})\Big\}\,,
\end{eqnarray}
where we have replaced $|\alpha|$ with $\sqrt{1-|\beta|^2}$ and ignored the phase of $\alpha$. A constant is dropped from the Hamiltonian. We rotate all qubits along some axis $\vec{n}$,
\begin{equation}\label{eqn:evolve_s}
    s^{x,y,z}_{a_j}(t)=\exp(2\pi i s^{\vec{n}}_{a_j}\frac{t}{T})s^{x,y,z}_{a_j} \exp(-2\pi i s^{\vec{n}}_{a_j}\frac{t}{T})\,,
\end{equation}
where $s^{\vec{n}}_{a_j}=\vec{n}\cdot \hat{s}=n_x s^{x}_{a_j}+n_y s^{y}_{a_j}+n_z s^{z}_{a_j}$ is the spin operator along the direction of $\vec{n}$ and $t\in [0,T]$. 
Thus at time $t$ the  Hamiltonian becomes
\begin{widetext}
\begin{eqnarray}\label{eqn:evolve_H}
    H(t)=&&\Delta \sum_{j=1}^m\Big\{ -s^z_{a_j}(t)s^z_{b_j}(t)-\frac{1}{2}(1-2|\beta|^2)[s^z_{a_j}(t)-s^z_{b_j}(t)] \nonumber \\
    &&+2\sqrt{1-|\beta|^2}[\mathrm{Re}(\beta)(s^x_{a_j}(t)s^x_{b_j}(t)+s^y_{a_j}(t)s^y_{b_j}(t))+\mathrm{Im}(\beta)(s^x_{a_j}(t)s^y_{b_j}(t)-s^y_{a_j}(t)s^x_{b_j}(t))]\Big\}~\,.
\end{eqnarray}
\end{widetext}
It is obvious that the eigen-energies of $H(t)$ do not change with $t$ and the corresponding eigenstates can be obtained by rotating those of $H_0$.  
Specifically, the energy gap $\delta(n)$ between the ground states and the first excited states
does not change with $t$. We are interested in the adiabatic rotation, where $T$ is big enough. In this case,  
according to Ref.\cite{Wilczek_Zee}, if the initial state $|\psi(0)\rangle$ lies in the subspace spanned by the degenerate ground states $\{|\psi_k(0)\rangle\}$, i.e. $|\psi(0)\rangle =\sum_{k} c_k |\psi_k(0)\rangle$, then the final state $|\psi(T)\rangle$ lies in the subspace spanned by the ground states $\{|\psi_k(T)\rangle\}$ as well.
Specifically, we have
\begin{equation}
    |\psi(T)\rangle=\sum_{kl} c_k U_{kl}|\psi_l(0)\rangle.
\end{equation}
where 
\begin{eqnarray}
    U=P[\exp(i\int_0^T \mathrm{d}t A(t))]\,, \\
    A_{kl}(t)= i\langle \psi_l(t) |\frac{\mathrm{d}}{\mathrm{d}t}|\psi_k(t)\rangle .
\end{eqnarray}
Here $A$ is the non-Abelian gauge matrix that drives the the dynamics in the subspace of the degenerate ground states. 
For the special case $\alpha=\beta=\sqrt{2}/2$, the gauge matrix has the following form 
\begin{equation}
    A_{kl}(t)=i\pi\langle \psi_l|\sum_{a=1}^n(s^+_a-s^-_a)|\psi_k\rangle~.
\end{equation}

Here is our algorithm. 
\begin{itemize}
    \item Choose a trivial solution of the Q2SAT problem as the initial state $|\psi(0)=|00...00\rangle$ and set $H(0)=H_0$. 
    \item Adiabatically rotate all qubits along some axis $\hat{n}$ from $t=0$ to $t=T$. During this rotation, the Pauli matrices  $s^{x,y,z}_{a_j}$
    of the qubit $a_j$ evolves  according to Eq. (\ref{eqn:evolve_s}) and the Hamiltonian of the system $H(t)$ evolves  according to Eq. (\ref{eqn:evolve_H}). 
    \item Make measurement at the end.
\end{itemize}

As is shown in \cite{farhi_goldstone_gutmann_sipser}, the time complexity of a quantum adiabatic algorithm is proportional to the inverse square of the energy gap $\delta(n)$ between the ground states and the first excited states. So, the time complexity of our algorithm depends how 
the energy gap $\delta(n)$ scales with $n$. Here we consider a special case to estimate the energy gap and examine how it is influenced by the coefficient $|\beta|$. 
In this special case, the spins form a one-dimensional chain and couple to their two neighbours. We assume that $\beta$ is a positive real number. 
The special case is in fact the well known Heisenberg chain and has been thoroughly studied. Its Hamiltonian is 
\begin{eqnarray}
    H_0 =&& \Delta \sum_{i}[-s^z_i s^z_{i+1}-\frac{1}{2}(1-2|\beta|^2)(s^z_i-s^z_{i+1})\nonumber\\
    &&+2|\beta| \sqrt{1-|\beta|^2}(s^x_i s^x_{i+1}+s^y_{i} s^y_{i+1})]~.
\end{eqnarray}
Its eigen-energies form a band $E_k$ and can be analytically found~\cite{Hodgson_Parkinson}. We examine two limits. 
When $\beta\rightarrow 0$, $E_k\rightarrow \Delta(1/2-2|\beta| \cos k )$, thus the gap approaches a constant, $\delta(n) \rightarrow \Delta/2$. 
When $\beta=\sqrt{2}/2$, $E_k=\Delta (1+\cos k)$, the gap $\delta=0$ if the chain is infinitely long .  
In our problem, due to that the chain has a finite length $n$, the wave vector $k$ is actually discrete and  we have $\delta(n)\sim O(n^{-2})$. As a result, We expect  that our algorithm to have the worst performance when $|\beta|$ approaches $\sqrt{2}/2$. According to the above analysis, we mainly investigate the performance of our algorithm at $\beta=\sqrt{2}/2$, and regard it as the worst performance.

\section{\label{sec:numerical_simulation}Numerical Simulation}
\begin{figure}
    \centering
    \includegraphics[height=5cm]{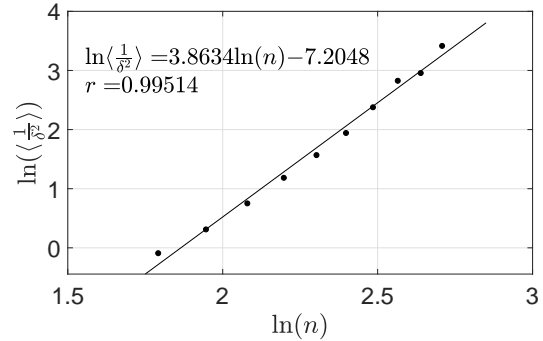}
    \caption{The relation between the average of the inverse square of energy gaps of randomly generated problems and the number of qubits. The x-axis is the logarithm of the number of qubits, $\log(n)$, and the y-axis is the logarithm of the average of the inverse square of the energy gap, $\log(\langle\frac{1}{\delta^2}\rangle)$. The line is fitted by least squares method.}
    \label{fig:gap}
\end{figure}

In our numerical simulation, we focus on the special case where  
\begin{equation}
    \ket{\Phi_j}=\frac{1}{\sqrt{2}} (|0_{a_j}1_{b_j}\rangle +|1_{a_j}0_{b_j}\rangle)\,.
\end{equation}
For this case, the Hamiltonian takes a simple form
\begin{equation}
    H_3 = \Delta \sum_{j=1}^m(s_{a_j}^{x} s_{b_j}^{x}+s_{a_j}^{y} s_{b_j}^{y}-s_{a_j}^{z} s_{b_j}^{z})~.
\end{equation}
This Hamiltonian commutes with the total angular momentum along the $z$ axis $[H_3,\sum_{a=1}^ns^z_a]=0$. 
The graph for each Hamiltionian is generated as follows. We first fix $n$, the number of vertices (or qubits), and then
generate edges between each pair of vertices with the probability $d$. As a result, the number of edges $m\approx dn(n-1)/2$. 
In our numerical calculation, we choose $d=0.1$. We randomly generate $10000$ graphs for $n=5$ to $n=11$, $1000$ 
for $n=12$, $n=13$ and $n=14$, and $100$ for $n=15$. 
The corresponding Hamiltonians are diagonalized numerically and the energy gap $\delta$ is extracted. 
The average of the energy gap $\delta$ is  plotted in logarithm scale in Fig.\ref{fig:gap}. Fitted by least squares method, we get
\begin{equation}
    \log(\langle\frac{1}{\delta^2}\rangle)=3.8634\log(n)-7.2048.
\end{equation}
with correlation coefficient $r=0.995$. 
This shows that the inverse square of the energy gap $\langle1/\delta^2\rangle\approx O(n^{3.9})$ and the time complexity $t\approx O(n^{3.9})$ according to Ref. \cite{farhi_goldstone_gutmann_sipser}. Such a time complexity is better than that of the quantum algorithm in \cite{Farhi_Kimmel_Temme}, which is of $O(n^{5.9})$ for $m\approx dn(n-1)/2$ and $1/\delta \approx O(n^{1.9})$. 

\begin{figure}
    \centering
    \includegraphics[height=5cm]{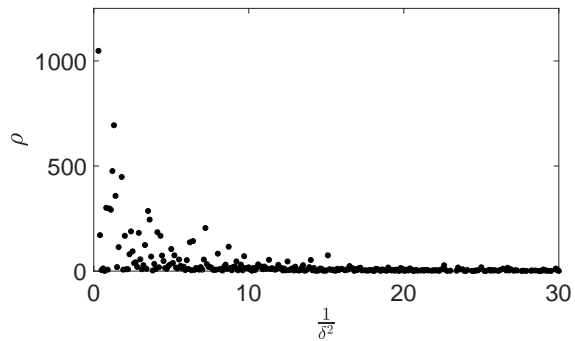}
    \caption{The distribution of the inverse square of energy gaps for $10000$ randomly generated problems with $n=11$. The x-axis is the inverse square of the energy gap, $\frac{1}{\delta^2}$, and the y-axis is the number $\rho$ of problems whose energy gap is within $[\frac{1}{\delta^2}-0.1, \frac{1}{\delta^2}]$. } 
    \label{fig:fig4}
\end{figure}

Shown in Fig.\ref{fig:fig4} is the distribution of the inverse square of energy gaps $1/\delta^2$ for a group of randomly generated graphs with $n=11$. The distribution shows that few graphs lead to a large inverse square of the gap, but most problems correspond to small inverse square of the gap near the average. Thus it is reasonable that we use the average of the inverse square of the energy gap to compute the time complexity. 

\begin{figure}
    \centering
    \subfigure{\includegraphics[height=5cm]{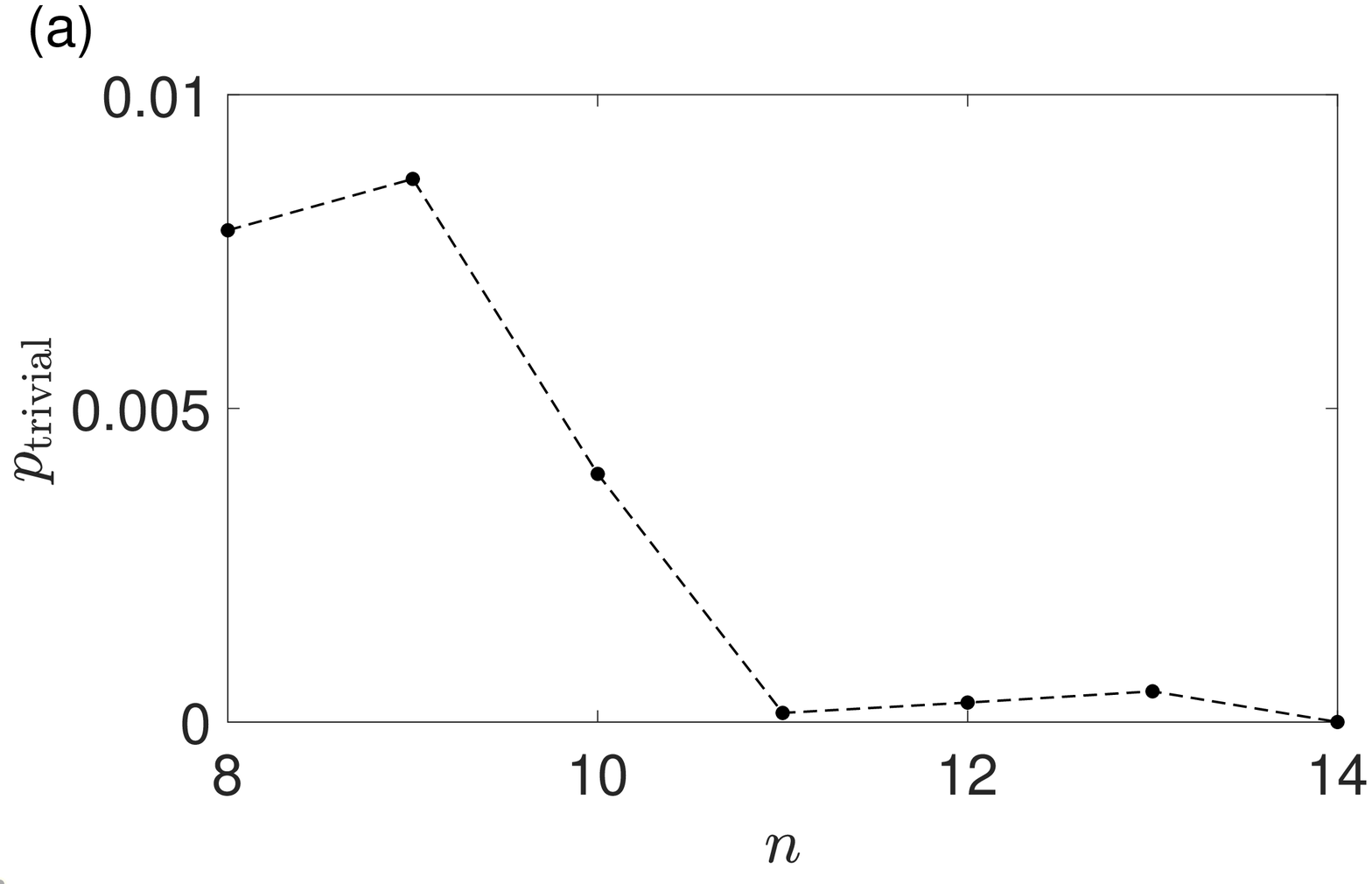}}
    \subfigure{\includegraphics[height=5cm]{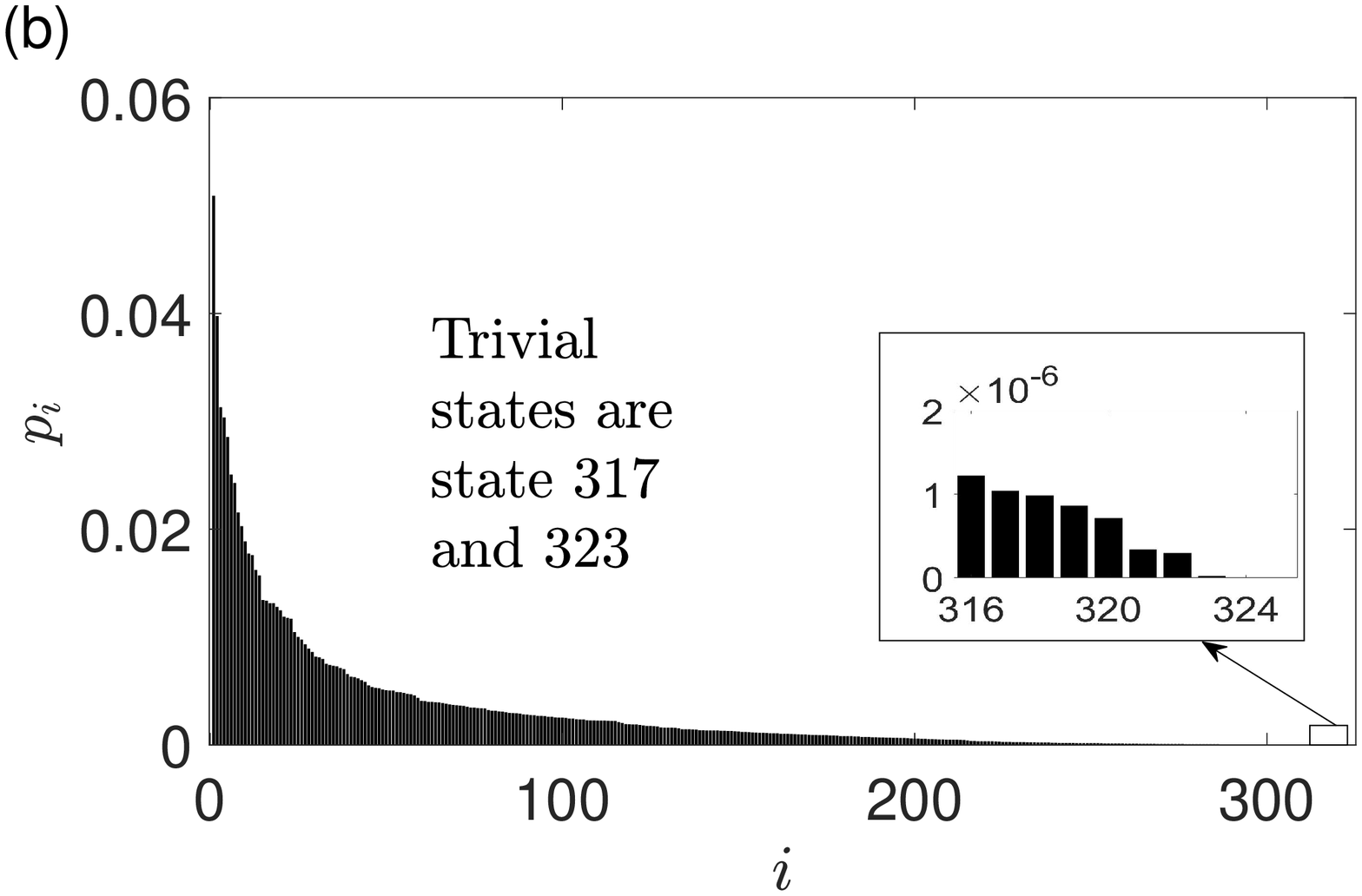}}
    \caption{(a) The probability of trivial states of a randomly generated Q2SAT problem after the adiabatic evolution for different $n$. The dashed line is guide for the eye. (b) The probability distribution of a randomly generated Q2SAT problem on all its ground states after the adiabatic evolution  for $n=14$. The x-axis is the index of all the ground states. The y-axis is the probabilities of ground states in the final state. The ground states are indexed according to their probabilities in the descending order. The index of the trivial states are 317 and 323. }
    \label{fig:fig1}
\end{figure}

Although our algorithm is quantum, we can still simulate it on our classical computer when the graph size is not very large.  
In our simulation,  we choose  the direction $\hat{n}$ to be along the $y$-axis. In this simple case, 
we have explicitly how the spin operators rotate 
\begin{equation}
    \left\{
    \begin{array}{lr}
        s^{x}_{a_j}(t)=s^{x}_{a_j}\cos(2\pi t/T)+ s^{z}_{a_j}\sin(2\pi t/T)\,,\\
        s^{y}_{a_j}(t)=s^{y}_{a_j}\,,\\
        s^{z}_{a_j}(t)=-s^{x}_{a_j}\sin(2\pi t/T)+ s^{z}_{a_j}\cos(2\pi t/T)\,.
    \end{array}
\right.
\end{equation}
In our numerics, we choose $T=\pi/(50\delta^2)$ for randomly-generated graphs with $n$ from 8 to 14. We simulate the evolution of the system by the fourth-order Runge-Kutta method, and calculate the module square of coefficients, i.e. probability, of the final state on all the possible ground states. The probability of the trivial states for graphs with $n$ from 8 to 10 and the probability distribution for a graph with $n=14$ are shown in Figure \ref{fig:fig1}. It can be seen that after the adiabatic evolution, we do not return to the trivial state but reach a non-trivial state with a high probability. Such an ability to find a non-trivial solution is better than that of the classical algorithm proposed by \cite{Beaudrap_Gharibian} and \cite{Arad_Santha_Sundaram_Zhang}. 

For our algorithm to work, a large number of solutions is required. However, when $d>0.1$, the number of solutions decreases significantly. In that case, the state remains on trivial states with a high probability after the adiabatic evolution, and our algorithm thus does not work any more. Numerical results show that for $n=14$ and $d>0.15$, our algorithm fails with a non-negligible probability.  

\section{\label{conclusion}Conclusion}
A quantum adiabatic algorithm for the Q2SAT problem is proposed. In the algorithm, the Hamiltonian is constructed so that  all the solutions of a Q2SAT problem are its ground states. A trivial product-state solution is chosen as the initial state. By rotating all the qubits, the system evolves adiabatically in the subspace of solutions and ends up on a non-trivial state. Theoretical analysis and numerical simulation
show that, for a set of Q2SAT problems,  our algorithm finds a non-trivial solution with time complexity better than the existing
algorithms.

\section{\label{acknowledgements}Acknowledgements}
The authors thank Tianyang Tao and Hongye Yu for useful discussions. This work is supported by the The National Key R\&D Program of China (Grants No.~2017YFA0303302, No.~2018YFA0305602), 
National Natural Science Foundation of China (Grant No. 11921005), and 
Shanghai Municipal Science and Technology Major Project (Grant No.2019SHZDZX01).

\providecommand{\noopsort}[1]{}\providecommand{\singleletter}[1]{#1}%

\end{document}